\documentclass[slac_one]{revtex4}
\usepackage{graphicx}
\usepackage{fancyhdr}
\def\d0{D\O}
\def\D0{D\O}

\newcommand{\Eslash}{\mbox{$E \kern-0.6em\slash$}}

\def\ifmath#1{\relax\ifmmode #1\else $#1$}%
\def\TeV{\ifmmode {\mathrm{ Te\kern -0.1em V}}\else
                   \textrm{Te\kern -0.1em V}\fi}%
\def\GeV{\ifmmode {\mathrm{ Ge\kern -0.1em V}}\else
                   \textrm{Ge\kern -0.1em V}\fi}%
\def\MeV{\ifmmode {\mathrm{ Me\kern -0.1em V}}\else
                   \textrm{Me\kern -0.1em V}\fi}%
\def\GeVcc{\ifmmode {\mathrm{ \GeV/c^2}}\else
                   \textrm{Ge\kern -0.1em V/c$^2$}\fi}%
\def\MeVcc{\ifmmode {\mathrm{ \MeV/c^2}}\else
                   \textrm{Me\kern -0.1em V/c$^2$}\fi}%

\def\Aslash{\mbox{${\hbox{$A$\kern-0.55em\hbox{/}}}$}}
\def\pslash{\mbox{${\hbox{$p$\kern-0.45em\hbox{/}}}$}}

\def\to{\rightarrow}

\def\gesim{\,{\raise-3pt\hbox{$\sim$}}\!\!\!\!\!{\raise2pt\hbox{$>$}}\,}
\def\lesim{\,{\raise-3pt\hbox{$\sim$}}\!\!\!\!\!{\raise2pt\hbox{$<$}}\,}
\def\boldoverdot{\,{\raise6pt\hbox{\bf.}\!\!\!\!\>}}

\def\diag{\hbox{\diag}}

\def\doubleundertext#1{
{\undertext{\vphantom{y}#1}}\par\nobreak\vskip-\the\baselineskip\vskip4pt%
\undertext{\hbox to 2in{}}}
\def\inbox#1{\vbox{\hrule\hbox{\vrule\kern5pt
     \vbox{\kern5pt#1\kern5pt}\kern5pt\vrule}\hrule}}
\def\sqr#1#2{{\vcenter{\hrule height.#2pt
      \hbox{\vrule width.#2pt height#1pt \kern#1pt
         \vrule width.#2pt}
      \hrule height.#2pt}}}
\def\today{\ifcase\month\or
  January\or February\or March\or April\or May\or June\or
  July\or August\or September\or October\or November\or December\fi
  \space\number\day, \number\year}
\def\pmb#1{\setbox0=\hbox{#1}%
  \kern-.025em\copy0\kern-\wd0
  \kern.05em\copy0\kern-\wd0
  \kern-.025em\raise.0433em\box0 }
\def\sumprime_#1{\setbox0=\hbox{$\scriptstyle{#1}$}
  \setbox2=\hbox{$\displaystyle{\sum}$}
  \setbox4=\hbox{${}'\mathsurround=0pt$}
  \dimen0=.5\wd0 \advance\dimen0 by-.5\wd2
  \ifdim\dimen0>0pt
  \ifdim\dimen0>\wd4 \kern\wd4 \else\kern\dimen0\fi\fi
\mathop{{\sum}'}_{\kern-\wd4 #1}}
\begin{document}

%Title of paper
\title{CP Violation Results from \d0} %% Paper title goes here

% Repeat the \author .. \affiliation  etc. as needed
%
% \affiliation command applies to all authors since the last
% \affiliation command. The \affiliation command should follow the
% other information

\author{John Ellison, for the \d0\ Collaboration}
\affiliation{University of California, Riverside, CA 92521, USA}
\author{}
\affiliation{}

%%%%%%%%%%%%%%% 4 pages is the limit

\begin{abstract}
We present results on CP violation from approximately $2.8$~fb$^{-1}$ of data collected by the \d0\ Experiment at the Fermilab Tevatron. The results presented are: 
(i) an improved measurement of the $B_s^0$ CP-violating phase from a flavor-tagged analysis of 
$B_s^0 \to J/\psi \; \phi$ decays;
(ii) a search for direct CP violation in $B^\pm \to J/\psi \; K^\pm (\pi^\pm)$ decays from a measurement of the
charge asymmetry $A_{CP}(B^\pm \to J/\psi \;  K^\pm)$; and
(iii) a search for indirect CP violation from searches for anomalous charge asymmetries in semileptonic $B_s^0$  decays.		  
\end{abstract}

%\maketitle must follow title, authors, abstract
\maketitle

\thispagestyle{fancy}

% body of paper here - Use proper section commands
% References should be done using the \cite, \ref, and \label commands
% Put \label in argument of \section for cross-referencing
%\section{\label{}}

\section{INTRODUCTION} % Section title should be in all capitals.

One of the primary goals of the Tevatron is the search for CP violation in $B$-meson decays. In this paper we describe some recent results in this area. The data used for the searches for CP violation described here are based on approximately $2.8$~fb$^{-1}$ of data collected by the \d0\ Experiment at the Fermilab Tevatron. The \d0\ detector consists of a central tracking system surrounded by a uranium liquid-argon calorimeter and an outer muon detection system~\cite{d0det}. The most important components of the detector for the measurements described here are the central tracker and the muon system. The central tracker has excellent coverage in the pseudorapidity range $|\eta| < 3$ and is immersed in a 2~T solenoidal magnetic field. A new silicon layer positioned close to the beam pipe was installed in 2006 and improves the impact parameter resolution. The muon system provides coverage and triggering out to  $|\eta| < 2$ and includes toroidal magnets for independent muon momentum measurement. During data taking, the polarity of the solenoidal and toroidal magnets were flipped every two weeks which allows systematics due to detector asymmetries to be reduced.

%%%\section{FLAVOR-TAGGED ANALYSIS OF \boldmath$B_s^0 \to J/\psi + \phi$ DECAYS}
\section{FLAVOR-TAGGED ANALYSIS OF $B_s^0 \to J/\psi + \phi$ DECAYS}

In the Standard Model, $B_s^0 - {\bar B}_s^0$ mixing is described by the Schr\"odinger Equation, 
\begin{eqnarray*}
i\frac{d}
{{dt}}\left( {\begin{array}{*{20}c}
   {\left| {B_s^0 } \right\rangle }  \\
   {\left| {\bar B_s^0 } \right\rangle }  \\

 \end{array} } \right) = \left( {\begin{array}{*{20}c}
   {M - i\frac{\Gamma }
{2}} & {M_{12}  - i\frac{{\Gamma _{12} }}
{2}}  \\
   {M_{12}^*  - i\frac{{\Gamma _{12}^* }}
{2}} & {M - i\frac{\Gamma }
{2}}  \\

 \end{array} } \right)\left( {\begin{array}{*{20}c}
   {\left| {B_s^0 } \right\rangle }  \\
   {\left| {\bar B_s^0 } \right\rangle }  \\

 \end{array} } \right)
\end{eqnarray*}

\noindent
The mass eigenstates are given by linear combinations of the weak eigenstates,
\begin{eqnarray*}
  \left| {B_L } \right\rangle  = p\left| {B_s^0 } \right\rangle  + q\left| {\bar B_s^0 } \right\rangle \\
  \left| {B_H } \right\rangle  = p\left| {B_s^0 } \right\rangle  - q\left| {\bar B_s^0 } \right\rangle
\end{eqnarray*}

\noindent
where $B_L$ denotes the light mass state (mostly CP even) and $B_H$ denotes the heavy mass state (mostly CP odd). CP violation occurs if $q/p \ne 1$ or if there is interference between direct and mixed decays of the $B_s^0$. In the SM the CP-violating phases are very small: the largest contribution comes form interference between direct and mixed decays and is given by~\cite{Lenz}
\begin{eqnarray*}
  \phi _s^{SM}  =  - 2\beta _s  =  - 2\arg \left( { - \frac{{V_{ts} V_{tb}^* }}{{V_{cs} V_{cb}^* }}} \right) \approx 0.04
\end{eqnarray*}

\begin{figure}
	\begin{center}
	\includegraphics[width=0.8\textwidth]{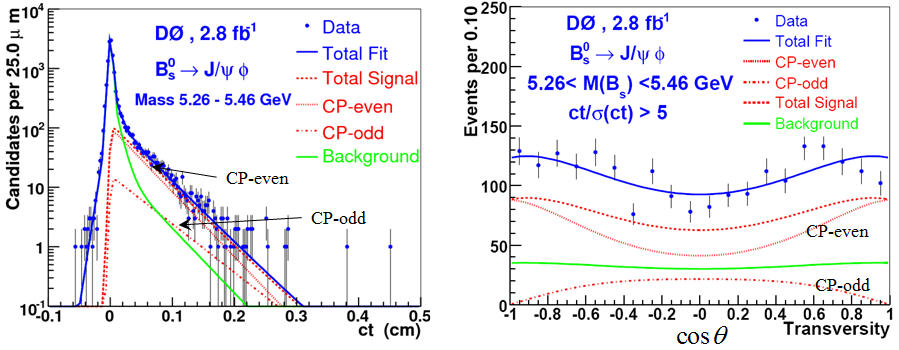}
	\caption{\label{fig:ct_trans_fit}
	Fits to the proper decay length (left) and transversity (right) distributions in $B_s \to J/\psi \phi$ events. The  fits show the relative contributions of the CP-even and CP-odd components, as well as the combinatorial background. }
	\end{center}
\end{figure}

\begin{figure}
\begin{tabular}{c c}
	\includegraphics[width=0.4\textwidth]{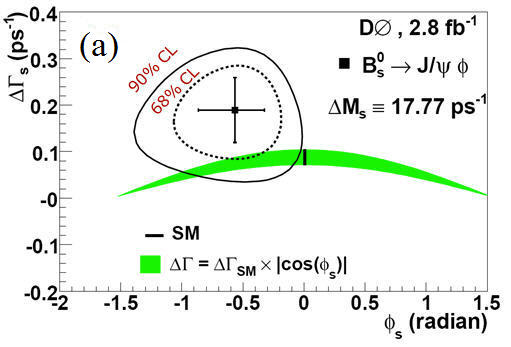} &
	\includegraphics[width=0.37\textwidth]{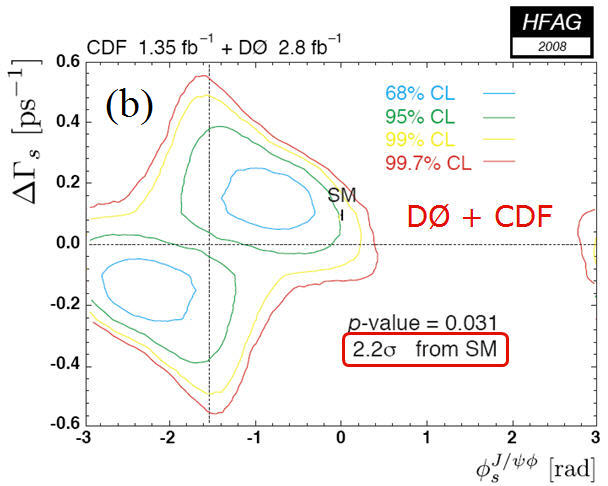}
\end{tabular}
	\caption{\label{fig:dg_phis}
	(a) Results of the fits in the $\phi_s - \Delta \Gamma_s$ plane for the \d0\ $B_s \to J/\psi \phi$ analysis based 
	on 2.8~fb$^{-1}$ of data. (b) Fit results from the combination of the \d0\ results with CDF results based on 
	1.35~fb$^{-1}$ of data.}
\end{figure}

The $B_s^0 \to J/\psi \; \phi$ decay involves a final state that is a mixture of CP-even and CP-odd states. In order to separate the CP-even and CP-odd states, we perform a maximum likelihood fit to the mass, lifetime, and time-dependent angular distributions of the $B_s^0 \to J/\psi (\to \mu^+ \mu^-) \; \phi (K^+ K^-)$ decay. The fit yields the CP-violating phase $\phi_s$ and the width difference $\Delta \Gamma_s \equiv \Gamma_L - \Gamma_H$. The decay can be described by three decay angles $\theta$, $\phi$, and $\psi$ defined in \cite{d0_phis_prl}. We employ initial state flavor tagging which improves the sensitivity to the CP-violating phase and removes a sign ambiguity on $\phi_s$ for a given $\Delta \Gamma_s$ present in our previous analysis~\cite{d0_phis_prl_previous}.  The projection of the fit results on the proper decay length and transversity ($\cos{\theta}$) distributions are shown in Fig.~\ref{fig:ct_trans_fit}. In the fit, $\Delta M_s$ is constrained to its measured value (from CDF) and the strong phases are constrained to values measured for $B_d$ at the $B$\textbf{}-factories, allowing some degree of violation of SU(3) symmetry. The $B_s$ flavor at production is determined using a combined opposite-side plus same­side tagging algorithm. Confidence level contours in the $\phi_s - \Delta \Gamma_s$ plane are shown in Fig.~\ref{fig:dg_phis}(a).

The fit yields a likelihood maximum at $\phi_s = -0.57^{+0.24}_{-0.30}$ and
$\Delta \Gamma_s = 0.19 \pm 0.07$~ps$^{-1}$, where the errors are statistical only.
As a result of the constraints on the strong phases, the second maximum, at  
$\phi_s = 2.92^{+0.30}_{-0.24}$, $\Delta \Gamma_s = -0.19 \pm 0.07$~ps$^{-1}$, is disfavored by a likelihood ratio of 1:29. 

From the fit results and studies of the systematic errors we obtain the width difference 
$\Delta \Gamma_s \equiv \Gamma_L - \Gamma_H = 0.19 \pm 0.07 {\rm (stat)}\thinspace ^{+0.02}_{-0.01} {\rm (syst)}$ ps$^{-1}$ and the CP-violating phase, $\phi_{s} = -0.57 ^{+0.24}_{-0.30} {\rm (stat)}\thinspace ^{+0.07}_{-0.02} {\rm (syst)}$.
The allowed 90\% C.L. intervals of $\Delta \Gamma_s$ and $\phi_s$ are $0.06 <\Delta \Gamma_s <0.30$ ps$^{-1}$ and $-1.20 <\phi_s < 0.06$, respectively. The probability to obtain a fitted value of $\phi_{s}$ lower than -0.57 given SM contributions only is 6.6\%, which corresponds to approximately 1.8~$\sigma$ from the SM prediction.

The results have been combined with CDF results by the Heavy Flavor Averaging Group. The CDF results are based on a data set of 1.35~fb$^{-1}$ and the combination was performed with no constraints on the strong phases. The results are shown in Fig.~\ref{fig:dg_phis}(b). The fit yields two solutions as follows:
\begin{eqnarray*}
  \phi_{s} &=& -2.37 ^{+0.38}_{-0.27}~\mathrm{rad}, ~~ \Delta \Gamma_s = -0.15 ^{+0.066}_{-0.059}~\mathrm{ps}^{-1} \\
  \phi_{s} &=& -0.75 ^{+0.27}_{-0.38}~\mathrm{rad}, ~~ \Delta \Gamma_s = 0.15 ^{+0.059}_{-0.066}~\mathrm{ps}^{-1} \\
\end{eqnarray*}
The $p$-value assuming only SM contributions is 3.1\%, corresponding to 2.2~$\sigma$ from the SM prediction.

\section{SEARCH FOR DIRECT CP VIOLATION IN {\boldmath$B^\pm \to J/\psi \; K^\pm (\pi^\pm)$} DECAYS}

\d0 has performed a search for direct CP violation by measurement of the charge asymmetry in $B^\pm \to J/\psi \; K^\pm (\pi^\pm)$ decays~\cite{d0_directCP}. The charge asymmetry is defined by

\begin{eqnarray*}
  A_{CP} (B^+   \to J/\psi \; K^+  (\pi^+  )) = 
  \frac{{N(B^-   \to J/\psi \;K^-  (\pi^-  )) - N(B^+   \to J/\psi \; K^+  (\pi^+  ))}}
  {{N(B^-   \to J/\psi \; K^-  (\pi^-  )) + N(B^+   \to J/\psi \; K^+  (\pi^+  ))}}
\end{eqnarray*}

\noindent
Direct CP violation in these decays leads to a non-zero charge asymmetry. In the SM there is a small level of CP violation due to the interference between tree-level and penguin decays and it is found that 
$A_{CP} (B^+   \to J/\psi \; K^+) \approx 0.003$~\cite{Hou} and
$A_{CP} (B^+   \to J/\psi \; \pi^+) \approx 0.01$~\cite{Dunietz}. New physics may significantly enhance $A_{CP}$. 

\begin{figure}
	\begin{center}
	\includegraphics[width=0.4\textwidth]{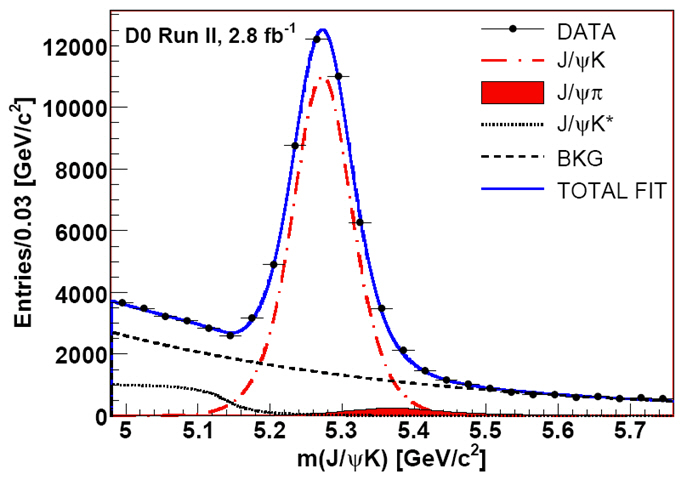}
	\caption{\label{fig:m_JpsiK}
	The $J/\psi K$ invariant mass distribution showing the result of the unbinned maximum likelihood fit.}
	\end{center}
\end{figure}

After selection of $B^\pm \to J/\psi \; K^\pm (\pi^\pm)$ candidates, the invariant mass of the $J/\psi K$ is constructed.
The mass distribution of the $J/\psi \; \pi$  system from the $B \to J/\psi \; \pi$ hypothesis is transformed
into the distribution of the $J/\psi K$ system by assigning the kaon mass to the pion. The $J/\psi K$ mass distribution is then fit to the sum of contributions from  $B^\pm \to J/\psi \; K^\pm$, $B^\pm \to J/\psi \; \pi^\pm$, $B^\pm \to J/\psi \; K^*$, and combinatorial background. From the fit we find approximately 40,000 $B^\pm \to J/\psi \; K^\pm$ candidates and about 1,600 $B^\pm \to J/\psi \; \pi^\pm$ candidates. Fig.~\ref{fig:m_JpsiK} shows the data and fit results.
To extract the charge asymmetry the sample is divided into 8 subsamples according to the solenoid polarity, the sign of the pseudorapidity of the $J/\psi K$ system, and the charge of the $K$ candidate. A $\chi^2$ fit to the number of events in each subsample yields the integrated raw charge asymmetry. This is then corrected for the asymmetry of the kaon interaction rate on nucleons to obtain the final results:
\begin{eqnarray*}
  A_{CP} (B^ +   \to J/\psi \;K^ +  ) &=&  + 0.0075 \pm 0.0061{\text{ (stat}}{\text{.) }} \pm 0.0027{\text{ (syst}}{\text{.)}} \\
  A_{CP} (B^ +   \to J/\psi \;\pi ^ +  ) &=&  - 0.09 \pm 0.08{\text{ (stat}}{\text{.) }} \pm 0.03{\text{ (syst}}{\text{.)}} \\ 
\end{eqnarray*} 
The results are consistent with the world average results~\cite{pdg}. The precision for $A_{CP} (B^ +   \to J/\psi \;\pi ^ +  )$ is comparable with the current world average, while for $A_{CP} (B^+   \to J/\psi \;K^ +  )$ the precision is a significant (factor of 2.5) improvement over the current world average.

%%%\section{SEARCH FOR DIRECT CP VIOLATION IN SEMILEPTONIC \boldmath$B_s$ DECAYS}
\section{SEARCH FOR DIRECT CP VIOLATION IN SEMILEPTONIC $B_s$ DECAYS}

\begin{figure}
	\begin{center}
	\includegraphics[width=0.4\textwidth]{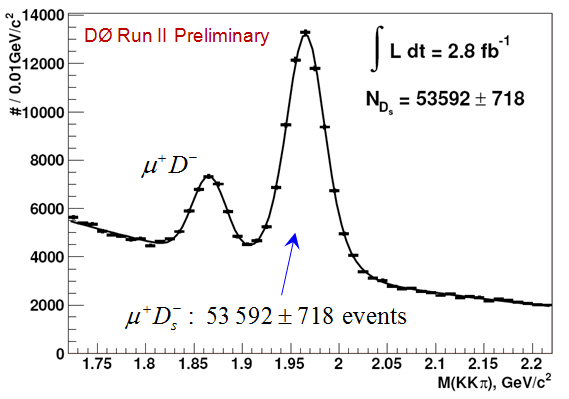}
	\caption{\label{fig:m_KKpi}
	The $KK \pi$ invariant mass distribution showing the $\mu^+ D^-$ and $\mu^+ D_s^-$ signals together with the fit results.}
	\end{center}
\end{figure}

In this section we report a new search for CP violation in the decay 
$B_s^0  \to D_s^-  \mu^+  \nu X, \;\; (D_s^- \to \phi \pi^- , \;\; \phi \to K^+ K^-)$
by measurement of the charge asymmetry using a time-dependent analysis with flavor tagging. The technique used is
similar to that used in the \d0\ $B_s$ oscillation analysis (see~\cite{d0_osc}). The flavor at production is determined from a combined opposite-side and same-side flavor tagging algorithm, while the flavor at decay is determined from the charge of the muon in the $B_s^0  \to D_s^-  \mu^+  \nu X$ decay. The proper decay length is then constructed from the measured visible proper decay length ($x^M$) and a $K$-factor: $c\tau _{B_s^0 }  = x^M K$, 
%%%,\quad {\text{where }}x^M  = \left[ {\frac{{\vec d_T^{B_s^0 }  \cdot \vec p_T^{\mu D_s^ -  } }}
%%%{{(p_T^{\mu D_s^ -  } )^2 }}} \right]cM_{B_s^0 }  
where the $K$-factor, defined by, $K = p_T^{\mu D_s^ -  } / p_T^{B_s^0 }$ is determined from Monte Carlo simulations.
A fit to the invariant $KK\pi$ mass of the selected data is shown in Fig.~\ref{fig:m_KKpi}.

The sample is divided into 8 subsamples according to solenoid polarity, the sign of the pseudorapidity of the $D_s \mu$ system, and the muon charge. An unbinned likelihood fit is used to extract the asymmetry. The systematic uncertainties are mainly due to uncertainties in the $c \bar c$ contribution, uncertainties in the efficiency vs visible proper decay length, and uncertainties in the $B_s \to D_s^{(*)} \mu \nu$ branching fractions. Accounting for these yields the final result: 
\begin{eqnarray*}
	a_{sl}^s  =  - 0.0024 \pm 0.0117{\text{ (stat}}{\text{.) }}_{ - 0.0024}^{ + 0.0015} {\text{ (syst}}{\text{.)}}
\end{eqnarray*}

This result is consistent with the SM prediction and is the most precise measurement to date.

\section{SUMMARY}

We have presented several new results from \d0\ on searches for CP violation in $B$-meson decays. A new measurement of the CP-violating phase in $B_s \to J/\psi \phi$ gives
\begin{eqnarray*}
	\phi_{s} = -0.57^{+0.24}_{-0.30} {\rm (stat)} \thinspace ^{+0.07}_{-0.02} {\rm (syst)}
\end{eqnarray*}

\noindent
The result has a $p$-value of 6.6\% for the SM-only hypothesis, or a significance of 1.8~$\sigma$. When combined with the CDF analysis, the $p$-value becomes 3.1\% which is equivalent to a 2.2~$\sigma$ deviation from the SM. Analysis of increased data sets is underway at \d0\ and CDF, and it will be very interesting to see if the deviation from the SM persists.

A search for direct CP violation was carried out using $B_s^\pm \to J/\psi \; K^\pm (\pi^\pm)$ decays. The measured charge asymmetries are consistent with the SM, but a significant improvement in precision by a factor 2.5 for the  
$B_s^\pm \to J/\psi \; K^\pm$ channel was achieved compared to the previous world average.

Finally, a new measurement of the charge asymmetry in semileptonic $B_s$ decay, reported for the first time at this conference, yields the most precise measurement of this quantity to date:  
$a_{sl}^s  =  - 0.0024 \pm 0.0117{\text{ (stat}}{\text{.) }}_{ - 0.0024}^{ + 0.0015} {\text{ (syst}}{\text{.)}}$.

As of the end of September 2008, \d0\ has recorded a total of 4.4~fb$^{-1}$ of data, compared with 2.8~fb$^{-1}$ used in the analyses reported here. Continued improvement of analysis techniques coupled with the much higher data sets expected in the next year or so at the Tevatron will provide much improved searches for CP violation in $B$-meson and $B_s$-meson decays.

% If you have acknowledgments, this puts in the proper section head.
%%%\begin{acknowledgments}
%%%The authors wish to thank JACoW for their guidance in preparing
%%%this template.

%%%Work supported by Department of Energy contract DE-AC02-76SF00515.
%%%\end{acknowledgments}

\end{document}